# Scattering by one-dimensional smooth potentials: between WKB and Born approximation

K.Yu. Bliokh[*,1,2], V.D. Freilikher[2], and N.M. Makarov[3]

[1]*Institute of Radio Astronomy, 4 Krasnoznamyonnaya st., Kharkov, 61002, Ukraine*

[2]*Department of Physics, Bar-Ilan University, Ramat Gan, 52900, Israel*

[3]*Instituto de Ciencias, Universidad Autónoma dePuebla, Puebla, Pue., 72050, México*

The paper discusses the applicability of WKB and Born (small perturbations) approximations in the problem of the backscattering of quantum particles and classical waves by one-dimensional smooth potentials with amplitudes small compared to the energy of the incident particle (above-barrier scattering). Both deterministic and random potentials are considered. The dependence of the reflection coefficient and localization length on the amplitude and the longitudinal scale of the scattering potential is investigated. It is shown that perturbation and WKB theories are inconsistent in the above-barrier backscattering problem. Not only the solutions but the regions of validity of both methods as well depend strongly on the details of the potential profile, and are individual for each potential. For deterministic potentials, a simple criterion that allows determining the boundary between the applicability domains of WKB and Born approximations is found.

PACS numbers: 03.65.Nk, 03.65.Sq, 42.25.Fx, 72.15.Rn

---

[*] E-mail: kostya@bliokh.kharkiv.com

## 1. Introduction

Two approximate methods are most often used in solving quantum mechanical and electrodynamical scattering problems. First one, the perturbation theory [1], is valid when the amplitude, $V_0$, of the scattering potential is small as compared to the energy, $E$, of the particle, so that a solution is sought as a series in powers of the small parameter

$$\delta = \frac{V_0}{E} \ll 1 \ . \tag{1}$$

The first term of this series is known as Born approximation.

The second method is WKB (quasiclassical) approximation [1–4], which is applied when the potential is a smooth function of coordinates, i.e. when the characteristic longitudinal scale of its variation, $L$, is large as compared to the characteristic wavelength $2\pi k_0^{-1} = 2\pi E^{-1/2}$, and the solution can be presented as an asymptotic expansion in powers of the small parameter

$$\varepsilon = \frac{1}{k_0 L} \ll 1 \ . \tag{2}$$

It may appear from Eqs. (1) and (2) that perturbation theory does not impose any restrictions on the dimensionless inverse scale $\varepsilon$, while WKB method is applicable for whatever small values of the dimensionless amplitude $\delta$ and therefore, when simultaneously $\delta \ll 1$ and $\varepsilon \ll 1$, both approximations should be valid and give the same result. It is known however [1,5–7], that if the longitudinal scale of the potential variations increases ($\varepsilon \to 0$) the convergence condition of the perturbation theory series is violated, no matter how small the fixed amplitude $\delta$ is. On the other hand, if parameter $\varepsilon$ is fixed ($\varepsilon \ll 1$) and the amplitude tends to zero ($\delta \to 0$), the WKB approximation breaks down [1,5–7]. (Note, that in papers [8–11] attempts have been made to construct the approximate theory including both Born and WKB theories. As shown in Refs. [5–7] they are actually incorrect in the WKB region.) This brings up the question: What are the regions of validity of WKB and small perturbation approximations when the scattering on a smooth ($\varepsilon \ll 1$) potential with small amplitude ($\delta \ll 1$) is concerned?

In the present paper we discuss the applicability of WKB and Born approximations for one-dimensional above-barrier scattering problems with different types of potentials, both deterministic and random, in the ballistic and localized regimes. It is shown that in the quasiclassical region, Eq. (2), the reflection coefficient, $R$, is extremely sensitive to the exact shape of the potential profile. When simultaneously $\varepsilon \ll 1$ and $\delta \ll 1$ there is no universal characteristic dependence $R(\delta, \varepsilon)$, and this function is quite individual for each given potential. This is in contrast to the case of tunneling ($\delta > 1$), when WKB approximation is robust in the sense that the transmission and reflection coefficients are determined by the characteristic height and width of the barrier and practically independent of the details of its shape. The surprising thing is that in the case of the above-barrier scattering even the regions of validity of WKB theory and Born approximation are essentially different for different potentials, so that universal inequalities restricting the applicability of the methods do not exist.

## 2. Basic equations

We consider the stationary one-dimensional Schrödinger equation:

$$\frac{d^2\psi}{dx^2} + [E - V(x)]\psi = 0 , \qquad (3)$$

where $E > 0$ is the energy (units $\hbar = 2m = 1$ are used), potential $V(x)$, is an analytical bounded function with a characteristic amplitude $V_0 = |V(x)|_{max}$ and a single characteristic longitudinal scale $L$ ($L^{-1} \sim |V'|/V_0$). By introducing variable $z = k_0 x$ and function $U(x/L) = U(\varepsilon z) \equiv V(x)/V_0$, we reduce Eq. (3) to the dimensionless form

$$\frac{d^2\psi}{dz^2} + [1 - \delta U(\varepsilon z)]\psi = 0 . \qquad (4)$$

Clearly, $U(\varepsilon z)$ is of order of unity, while its derivative with respect to $z$ is of order of $\varepsilon$. In what follows we consider the above-barrier scattering in the sense that $\delta < 1$ (more precisely, the inequality $\delta\varepsilon/(1-\delta) \ll 1$ is necessary in order WKB approximation to be valid for all real $z$ [5]).

In the Born approximation the reflection coefficient for a quantum particle (or classical wave) is given by [1]:

$$R_{Born} = \frac{\delta^2}{4}\left|\int_{-\infty}^{\infty} U(\varepsilon z)e^{2iz}dz\right|^2. \tag{5}$$

Eq. (5) represents the first term of the perturbation series. When the amplitude of $2k_0$-harmonic (corresponding to the first resonant Bragg backscattering) in the power spectrum of the potential is zero, higher terms of the perturbation expansion should be calculated.

In the quasiclassical limit, Eq. (2), the reflection coefficient can be calculated by means of the standard WKB procedure. Taking into account that in the case under consideration ($\delta < 1$) all turning points, i.e. the solutions of the equation $1 - \delta U(\varepsilon z) = 0$, are located in the complex $z$-plane, and main contribution to the reflection coefficient is given by the closest to the real axis complex turning point $z_0$ in the upper half-plane, one obtains [1–4]:

$$R_{WKB} = \exp\left(-4\,\mathrm{Im}\int_{z_r}^{z_0}\sqrt{1-\delta U(\varepsilon z)}\,dz\right), \tag{6}$$

where $z_r$ is an arbitrary point on the real $z$ axis.

Regions of applicability of Eqs. (5) and (6) are well established when only one of the parameters, $\varepsilon$ or $\delta$, is small, and can be written respectively [1]

$$\frac{\delta}{\varepsilon} \ll 1, \quad (\varepsilon \geq 1); \tag{7a}$$

$$\varepsilon \ll 1, \quad (\delta \sim 1). \tag{7b}$$

When $\delta \ll 1$ and $\varepsilon \ll 1$ simultaneously the situation is much more complicated, Eqs. (7) are irrelevant, and moreover, universal conditions for the applicability of both perturbation theory and WKB approximation do not exist. Below this problem is investigated for the deterministic (Sec. 3) and random scattering potentials (Sec. 4).

Note, that smallness of the amplitude $\delta$ gives rise to violation of the WKB approximation Eq. (6), since in this case the turning point $z_0$ is located far from the real axis and, as can be seen, it nears the singular point $z_1$ of the potential: $U(\varepsilon z_1) = \infty$. Indeed, $\delta U(\varepsilon z_0) = 1$, and $U(\varepsilon z_0) \to \infty$ at $\delta \to 0$. Therefore the singularity can make a comparable contribution to the backscattering [1,5–7].

## 3. Backscattering by deterministic barriers

We consider the applicability domains of WKB and Born approximations, and dependences $R(\delta, \varepsilon)$ with simple examples of the analytical potential $U(\varepsilon z)$.

**Example I** (Fig. 1a):

$$U(\varepsilon z) = \frac{1}{1 + e^{-\varepsilon z}} \ . \tag{8}$$

When the energy of the particle is larger than the maximum height of the potential ($\delta < 1$, above-barrier scattering) the exact solution for the reflection coefficient is given by [1]:

$$R^{(I)} = \left[ \frac{\operatorname{sh} \pi \varepsilon^{-1}\left(1 - \sqrt{1-\delta}\right)}{\operatorname{sh} \pi \varepsilon^{-1}\left(1 + \sqrt{1-\delta}\right)} \right]^2 \ . \tag{9}$$

If condition Eq. (1) holds, the Taylor expansion of Eq. (9) in powers of $\delta \ll 1$ gives

$$R^{(I)} \approx \left(\frac{\pi \delta}{\varepsilon}\right)^2 e^{-\frac{4\pi}{\varepsilon}} \quad \text{at} \quad \frac{\delta}{\varepsilon} \ll 1 \ , \tag{10a}$$

that exactly corresponds to the Born approximation Eq. (5). In the limit Eq. (2) the expansion of Eq. (9) yields

$$R^{(I)} \approx e^{-\frac{4\pi}{\varepsilon}} \quad \text{at} \quad \frac{\delta}{\varepsilon} \gg 1 \ , \tag{10b}$$

that coincides precisely with the WKB asymptotic given by Eq. (6) after the integral there is calculated.

Thus, for the given potential, Eq. (8), the boundary between the domains of applicability of WKB and Born approximations lies at $\delta/\varepsilon \sim 1$ where both results have the same order of magnitude.

Since Eq. (6) takes into account only the contribution of the simple complex turning point $\varepsilon z_0 = i\pi + \ln(1-\delta)^{-1}$ and neglects the effect of the first-order pole $\varepsilon z_1 = i\pi$, the region of validity of the WKB approximation is restricted by the condition that these points should not be too close to each other,

$$|z_0 - z_1| \gg 1 . \qquad (11)$$

Easy to show, the condition Eq. (11) coincides with the estimation of the correction term in Eq. (10b). The transition between WKB and Born asymptotics for the case of simple turning point and the first-order pole was discussed in Refs. [5–7].

**Example II** (Fig. 1b):

$$U(\varepsilon z) = \frac{1}{\operatorname{ch}^2 \varepsilon z} . \qquad (12)$$

For this potential the exact expression for the above-barrier reflection coefficient can also be obtained analytically [1]:

$$R^{(II)} = \frac{\operatorname{ch}^2\left(\frac{\pi}{2}\sqrt{\frac{8\delta}{\varepsilon^2}-1}\right)}{\operatorname{sh}^2\frac{\pi}{\varepsilon} + \operatorname{ch}^2\left(\frac{\pi}{2}\sqrt{\frac{8\delta}{\varepsilon^2}-1}\right)} . \qquad (13)$$

The expansion of this expression in powers of $\delta \ll 1$ yields:

$$R^{(II)} \approx \frac{4\pi^2 \delta^2}{\varepsilon^4} e^{-\frac{2\pi}{\varepsilon}} \quad \text{at} \quad \frac{\delta}{\varepsilon^2} \ll 1 , \qquad (14a)$$

that coincides with the Born approximation given by the general formula Eq. (5). The expansion of Eq. (13) with $\varepsilon \ll 1$ gives

$$R^{(II)} \approx e^{-\frac{2\pi}{\varepsilon}} \quad \text{at} \quad \frac{\delta}{\varepsilon^2} >> 1 \; . \tag{14b}$$

The identical formula is given by the WKB expression Eq. (6).

In distinction to Example I, the dividing line between the applicability domains of Born and WKB approximations, on which two asymptotics (14) are of the same order of magnitude lies at $\delta / \varepsilon^2 \sim 1$.

Equation (4) with the potential Eq. (12) has two turning points, $\varepsilon z_0 = i(\pi/2 \pm \sqrt{\delta})$, and the closest to them second-order pole $\varepsilon z_1 = i\pi/2$. The contribution of the singularity is negligible, i.e. Eq. (6) is valid, if $|z_0 - z_1| >> 1$, which gives the condition of Eq. (14b).

**Example III** (Fig. 1c):

$$U(\varepsilon z) = e^{-(\varepsilon z)^2} \; . \tag{15}$$

In this case explicit analytical formula for the reflection coefficient is unknown, and the asymptotics for $\delta << 1$, $\varepsilon << 1$ and the corresponding correction terms cannot be obtained as limiting cases of an exact analytical result, as it was done in previous examples. However, general formulas for Born (Eq. (5)) and WKB (Eq. (6)) approximations are valid, and upon substitution Eq. (15) yield respectively

$$R_{Born}^{III} = \frac{\pi \delta^2}{\varepsilon^2} e^{-\frac{2}{\varepsilon^2}} , \tag{16a}$$

$$R_{WKB}^{III} = \exp\left[-4\int_0^{z_0}\sqrt{1-\delta e^{-(\varepsilon z)^2}}\,dz\right] \sim e^{-\frac{1}{\varepsilon}\sqrt{\ln\delta^{-1}}} , \tag{16b}$$

where $\varepsilon z_0 = i\sqrt{\ln \delta^{-1}}$ is a simple turning point.

Let us find the applicability conditions for Eqs. (16a) and (16b). Straightforward evaluation of the region of applicability of Eq. (16a) by calculating the second-order term in the Born series is rather stubborn and cumbersome procedure. The condition Eq. (11) used in examples I and II is not applicable now, since the singularity of the potential Eq. (15) lies at infinity: $z_1 = i\infty$. But one can find the limits of validity of WKB result Eq. (16b) in a relatively simple way. Indeed, WKB method is based on the assumption that the characteristic scale of the spatial variations of the solution for Eq. (4) (i.e. the wavelength that is of order of one in our dimensionless units) is much smaller than that of the scattering potential. As complex variable $z$ approaches a singularity of the potential, the scale of potential variation decreases. In the contour of integration in Eq. (6) the turning point $z_0$ is the closest to the singularity $z_1$. Then the condition of WKB approximation applicability can be formulated as a requirement of smoothness of the potential at the turning point:

$$\left.\frac{|U(\varepsilon z)|}{|U'(\varepsilon z)|}\right|_{z=z_0} = \frac{1}{\delta |U'(\varepsilon z)|_{z=z_0}} \gg 1 \ . \tag{17}$$

For the potential Eq. (15) this condition takes the form:

$$\delta e^{\varepsilon^{-2}} \gg 1 \ . \tag{18}$$

One can see that both results, Eqs. (16), are of the same order (with accuracy of estimate (16b)) when $\delta e^{\varepsilon^{-2}} \sim 1$. This fact, together with Eq. (18), leads to conclude that the validity of small perturbation approximation, Eq. (16a), is restricted by the inverse to Eq. (18) inequality $\delta e^{\varepsilon^{-2}} \ll 1$.

The inequality Eq. (17) is rather general condition of the applicability of WKB approximation in above-barrier scattering problems. Easy to show that Eq. (11) derived in two previous examples (and in Refs. [5–7]) follows from Eq. (17) as a particular case. Actually, this condition can be regarded as a simple criterion that allows determining the boundary between the applicability domains of WKB and Born approximations.

The foregoing examples clearly demonstrate that when the scattering potential is weak and smooth ($\delta \ll 1$ and $\varepsilon \ll 1$), regions of applicability of WKB and Born approximations do not overlap and are separated by a certain line $\alpha(\delta,\varepsilon) = 1$ in $(\delta,\varepsilon)$ domain. At this line both approximations result in the reflection coefficient of the same order. A universal function $\alpha(\delta,\varepsilon)$ does not exist and the explicit form of $\alpha(\delta,\varepsilon)$ (as well as of the dependences $R(\delta,\varepsilon)$) is crucially determined by the explicit form of the potential.

## 4. Localization in smooth random potentials

It is known (see, for example, Ref. [12]) that weak (above-barrier) scattering by a one-dimensional random potential of a length $L_0$ (as previously, all lengths are dimensionless and measured in the units of the wavelength $k_0^{-1}$) leads to localization effects if $L_0$ is sufficiently large. It means that the transmission coefficient at typical (most probable) realizations is exponentially small: $T_{typ} \sim e^{-2L_0/l_{loc}}$, $(L_0 \gg l_{loc})$. Here $l_{loc}$ is so-called localization length defined as $l_{loc}^{-1} = -(2L_0)^{-1} \langle \ln T \rangle$ ($\langle \ldots \rangle$ means averaging over the ensemble of random realizations of the potential $U(\varepsilon z)$). In the weak scattering limit it can be calculated in the small perturbation approximation and has the form [12]

$$l_{loc}^{-1} = \frac{\delta^2 w(2)}{4}, \qquad (19)$$

were $w(2) = \int \widetilde{w}(z) e^{2iz} dz$, and $\widetilde{w}(z)$ is the binary correlation function of the potential $U(\varepsilon z)$. From Eq. (19) it follows that in the weak scattering approximation $l_{loc}^{-1}$ is always proportional to $\delta^2$, whereas its dependence on parameter $\varepsilon$ essentially determined by the form of binary correlation function $\widetilde{w}(z)$ and can be different for different potentials $U(\varepsilon z)$.

As it was shown above, at $\varepsilon \to 0$ (and fixed $\delta \ll 1$) the perturbation theory ceases to be true, and the problem can be solved in WKB approximation. Backscattering on smooth random potentials in WKB limit was examined in Refs. [4,13,14]. It was shown that effectively the above-barrier scattering occurs in the vicinity of the complex turning points, which in the case of random potential are randomly distributed in complex $z$-plane. Typical distance (along $\operatorname{Re} z$ axis) between turning points is of order of $\varepsilon^{-1} \gg 1$, and therefore single scattering acts at different turning points can be considered as statistically independent. Thus, the problem reduces to the statistical averaging of the reflection coefficient of a single turning point, Eq. (6), over the distribution of all (random) turning points, which yields [14]:

$$l_{loc}^{-1} \sim \int_0^\infty M(\xi) e^{-4\xi} d\xi \ . \tag{20}$$

Here $M(\gamma)$ is the distribution function of the random exponent $\gamma = \text{Im} \int_{z_r}^{z_0} \sqrt{1 - \delta U(\varepsilon z)} dz$ in Eq. (6), or in other words, is the average number of turning points in the range $(\gamma, \gamma + d\gamma)$ and in the unit interval $\text{Re} \, z$.

If the distribution function $M(\gamma)$ has a sufficiently sharp maximum at certain $\gamma = \gamma_{max}$, the integral Eq. (20) can be roughly estimated as

$$l_{loc}^{-1} \sim \lambda^{-1} e^{-\gamma_{max}} \ , \tag{21a}$$

where $\lambda$ is an average distance along axis $\text{Re} \, z$ between turning points with $\gamma \sim \gamma_{max}$. Similar formula was derived in Refs. [4,13]. Since we consider potentials with a single longitudinal characteristic scale, it is natural to assume that $\lambda \sim \varepsilon^{-1}$. Besides, if $\delta \sim 1$, then $\gamma_{max} \sim \varepsilon^{-1}$, and Eq. (21a) leads to

$$l_{loc}^{-1} \sim \varepsilon e^{-\varepsilon^{-1}} \ . \tag{21b}$$

Eqs. (20), (21) can, in principle, be used to calculate (or at least to estimate) the localization length, provided the parameters of the system under consideration are in the range of validity of WKB approximation. However, as is follows from the results of Sec. 3, finding of this range when $\delta \ll 1$ and $\varepsilon \ll 1$ presents substantial difficulties.

Indeed, propagation over any random realization of potential may be conceived as successive scatterings on elementary barriers of different shapes (like those considered in Sec. 3.). Each elementary reflection coefficient should be calculated with Eq. (5) or Eq. (6), depending on which side of the dividing line, $\alpha_i(\delta, \varepsilon) = 1$, parameters $\delta$ and $\varepsilon$ are located.

Obviously, all these lines will fill out a certain area in $(\delta, \varepsilon)$ plane (see Fig. 2) where both approximations are inapplicable, and therefore Eqs. (20), (21) are invalid.

As an example of unjustified use of WKB approximation that may produce a questionable result we cite the formula for the inverse localization length obtained in Ref. [14] for the potential $U(\varepsilon z) = m^2(\varepsilon z)$ and small amplitude $\delta$, where $m(\varepsilon z)$ is the Gaussian process with zero average:

$$l_{loc}^{-1} \sim \varepsilon^{-1/3} \delta^{-1/6} \exp\left(-\varepsilon^{-2/3} \delta^{-1/3}\right). \qquad (22)$$

Note that the fractional dependence on the inverse scale $\varepsilon$ in Eq. (22) is rather unusual for the WKB theory. It may be connected with following circumstances. First, only the contribution of the closest to real axis turning points with maximum derivates $m'$ (minimal $\gamma$) was taken into account in Eq. (20). However maximum values of the Gaussian random quantity $m'$ can be (with finite probability) large and violate the condition of applicability of WKB approximation Eq. (17) at any finite $\varepsilon$. Second, main contribution to the integral Eq. (20) can be made not by the turning points with minimal $\gamma$. Indeed, if there exist a sharp enough maximum $M(\gamma)$, the integral Eq. (20) gives the estimates Eqs. (21). And finally, it can be said with confidence that Eq. (22) is incorrect at sufficiently small $\delta$, where the perturbation theory is applicable and $l_{loc}^{-1} \propto \delta^2$.

## 5. Discussion

Comparison analysis of WKB and perturbation methods carried out in Sec. 3 has demonstrated that, as applied to the calculation of the reflection coefficient for weak ($\delta \ll 1$) and smooth ($\varepsilon \ll 1$) potential barriers, the regions of validity of these approximations do not overlap. In $(\delta, \varepsilon)$ plane they are separated by a line, $\alpha(\delta, \varepsilon) = 1$, on which both approximations match, i.e. give the same order of magnitude of the reflection coefficient. The explicit shape and location of this dividing line, as well as the explicit form of the function $R(\delta,\varepsilon)$ depend drastically on the shape of the scattering potential (Fig. 2). Recall that when only one of parameters, $\delta$ or $\varepsilon$, is small the conditions of applicability are insensitive to the details of the potential: if $\varepsilon \geq 1$ perturbation theory is valid for $\delta/\varepsilon \ll 1$, in the case $\delta \sim 1$ WKB approximation applies when $\varepsilon \ll 1$.

From mathematical point of view, WKB and perturbation approximations do not overlap when $\delta \ll 1$ and $\varepsilon \ll 1$ since the first one describes the asymptotic behaviour of the reflection coefficient at $\varepsilon \to 0$, while the second one presents the series of Taylor type. Indeed, the perturbation series has a finite convergence radius $\delta_0 > 0$. This radius is a function of the inverse scale $\varepsilon$: $\delta_0 = \delta_0(\varepsilon)$, and $\delta_0(\varepsilon) \to 0$ at $\varepsilon \to 0$ (see Eqs. (7a), (10a), (14a), (18)). Since at $\varepsilon \to 0$ the convergence radius of the perturbation series vanishes, WKB theory for above-barrier backscattering is never consistent with Born approximation.

To get physical insight into the incompatibility of two methods, let us recall that small perturbation theory describes the above-barrier reflection as a series of multiple resonant Bragg scatterings by different Fourier harmonics of the potential. This series converges, and Born approximation gives correct result when the reflection is determined by single resonant backscattering (first Bragg resonance) from the $2k_0$ spectral Fourier component of the potential (see Eq. (5)). In this case contributions of multiple scatterings by harmonics with larger periods (higher resonances) are not essential. However, if the potential is smooth ($\varepsilon \to 0$) so that WKB theory applies, large-scale harmonics corresponding to the higher-order resonances are predominant in the spectrum of the potential, and determine the scattering pattern.

Thus, small perturbation and quasiclassical methods are related to two limiting cases when the reflection coefficient is determined by the first and high-order Bragg resonances respectively, and therefore have different (non-overlapping) regions of applicability. In examples of regular deterministic potentials considered in Sec. 3 these regions match at a dividing line in $(\delta,\varepsilon)$ plane. If a random potential is concerned, the line turns into an area in $(\delta, \varepsilon)$ domain where neither of two approximations is valid. But even within the range of applicability of WKB approach practical utilization of the seemingly simple formula Eq. (20) is rather problematic because it contains the distribution function of the turning points, $M(\gamma)$, which is usually unknown. Moreover, it is unclear whether an unambiguous correspondence between statistics of the potential and that of its turning points exists in principle. In any case, the statistics of turning points of the random potential (and therefore the localization length in the quasiclassical regime) is substantially determined by all higher moments, unlike the Born approximation for which the knowledge of the binary correlation function is sufficient.

Demonstrated above high sensitivity of the quasiclassical reflection coefficient to the details of the scattering potential is not of theoretical interest only. It must be taken into account in processing of scattering data, if one wants to compare them with the corresponding theoretical results. In real experiments, the shape of the scattering potential (which is an input in the WKB calculations) is known, at best, in a set of discrete points, but most commonly only the characteristic scales, $\delta$ and $\varepsilon$ are available. While it is sufficient to estimate the tunnelling transmission coefficient (if $\delta > 1$), when the above-barrier reflection is concerned, a small difference in the shape of the fitting function can lead to a dramatic difference in the predicted value of $R$. A vivid example is furnished by Fig. 2. Profiles Fig. 1b and Fig. 1c are practically undistinguishable; nonetheless the corresponding refection coefficients have little in common.

To conclude, a general criterion of the applicability of WKB approximation (Eq. (17)), has been formulated. Applicability of perturbation and WKB approximations to calculations of the reflection coefficient from deterministic and random potential profiles has been studied. When only the dimensionless amplitude of the potential is small ($\delta \ll 1$, $\varepsilon > 1$), perturbation theory is valid, and the reflection coefficient (in the ballistic regime) and the inverse localization length (in the localization regime) are universally proportional to $\delta^2$, while the dependence on inverse scale $\varepsilon$ is individual for each potential. If only the inverse scale of the potential is small ($\varepsilon \ll 1$, $\delta \sim 1$) WKB theory applies predicting universal $\varepsilon$-dependences $R(\varepsilon) \sim e^{-1/\varepsilon}$ and $l_{loc}^{-1}(\varepsilon) \sim \varepsilon e^{-1/\varepsilon}$. There are no universal dependences of $R$ and $l_{loc}$ upon amplitude $\delta$ in this case. In the event of the above-barrier scattering by smooth potential profiles, when $\delta \ll 1$ and $\varepsilon \ll 1$ simultaneously, the regions of validity of two theories do not overlap. Not only the explicit form of function $R(\delta, \varepsilon)$ but also the location and the shape of the line that separates (in $(\delta, \varepsilon)$ plane) the applicability domains of two theories depend drastically on the explicit form of the potential. and $l_{loc}^{-1}(\delta, \varepsilon)$. When it comes to calculating the localization length, the line turns into a finite area where both approaches can be invalid. Apparently the size of this region in $(\delta, \varepsilon)$ plane depends on the statistics of the potential.

**Figure captions**

**Fig. 1.** Potentials $U(\varepsilon z)$. Figs. (a)–(c) depict the potentials of the examples I–III, Sec. 3, respectively.

**Fig. 2.** Dividing lines separating the domains of applicability of WKB and perturbation theories, for different deterministic potentials on $(\ln \varepsilon^{-1}, \ln \delta^{-1})$ plane. Curves (a)–(c) correspond to the potentials of examples I–III, Sec. 3. They are determined by the conditions from Eqs. (10), (14) and (18), and satisfy the equations $\ln \delta^{-1} = \ln \varepsilon^{-1}$, $\ln \delta^{-1} = 2\ln \varepsilon^{-1}$ and $\ln \delta^{-1} = e^{2\ln \varepsilon^{-1}}$, respectively. The domains of applicability of Born approximation lie on the left from the corresponding curves, while that of WKB approximation lie on the right. For random potentials in the area between the lines neither perturbation nor WKB theory is applicable.

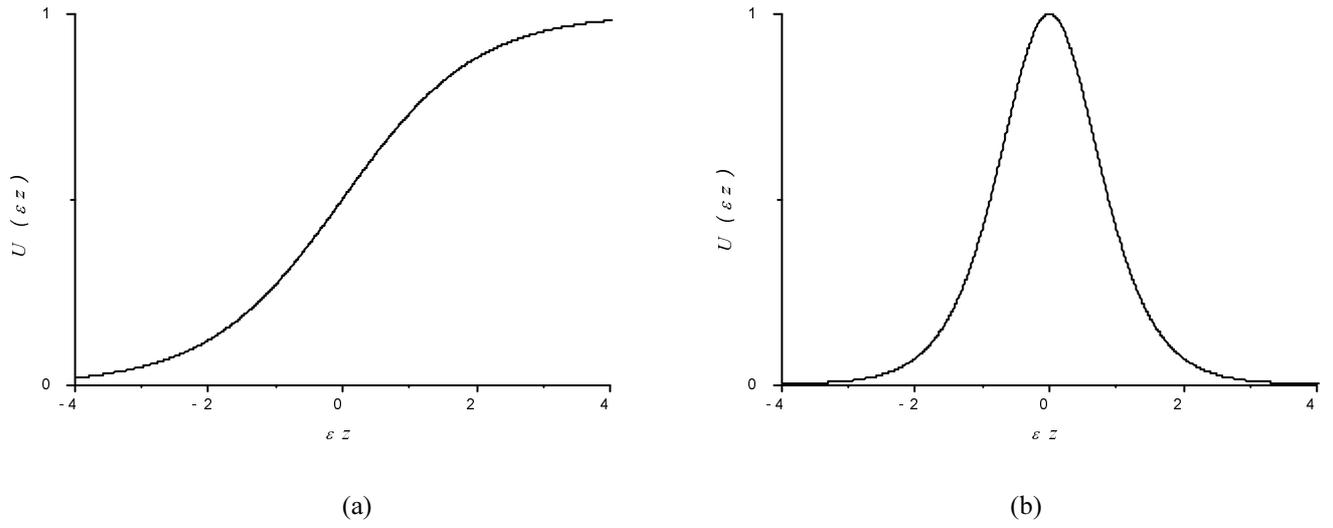

(a)

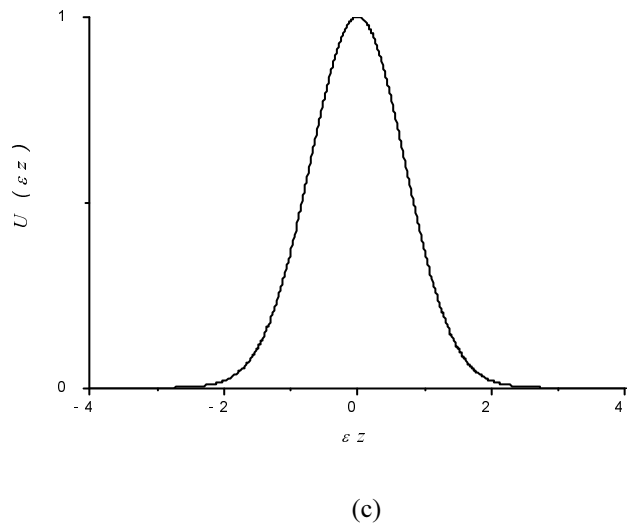

(b)

(c)

**Fig. 1**

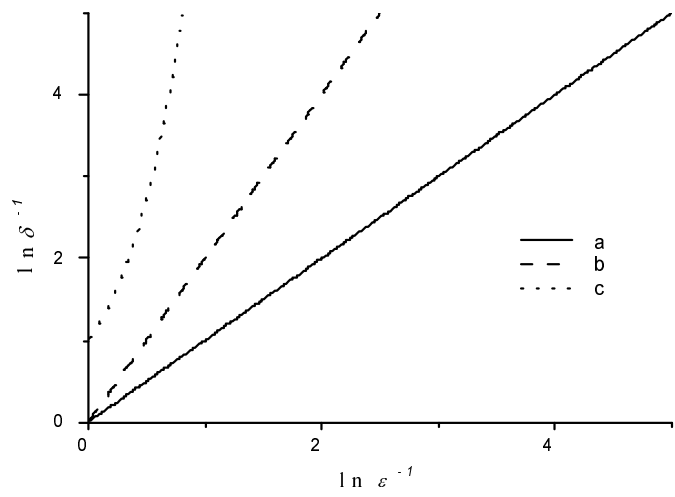

**Fig. 2**